\documentclass[preprint]{aastex}

\usepackage{epsfig}

\def\gx{GX~339$-$4}
\def\cx{Cyg~X$-$1}

\def\gro{GRO~J1655$-$40}
\def\grsds{GRS~1758$-$258}

\def\1e{1E~1740.7$-$2942}

\def\xte{XTE~J1550$-$564}

\slugcomment{Accepted for publication in ApJ}

\shorttitle{Corbel et al.}
\shortauthors{X-ray states and radio emission in \xte}

\begin{document}


\title{X-ray states and radio emission in the black hole
candidate \xte }


\author{S.~Corbel, P.~Kaaret}
\affil{Harvard--Smithsonian Center for Astrophysics, 60 Garden Street, Cambridge, MA 02138}
\email{corbel@head-cfa.harvard.edu}

\author{R.K.~Jain, C.D.~Bailyn}
\affil{Yale University, Department of Astronomy, P.O. Box 208101, New Haven, CT 06520-8101}

\author{R.P.~Fender}
\affil{Astronomical Institute `Anton Pannekoek', University of Amsterdam, 
and Center for High Energy Astrophysics, Kruislaan 403, 1098 SJ Amsterdam, 
The Netherlands }

\author{J.A.~Tomsick, E.~Kalemci}
\affil{Center for Astrophysics and Space Sciences, University of California, San Diego, MS 0424, La Jolla, CA 92093}

\author{V.~McIntyre}
\affil{Australia Telescope National Facility, PO Box 76, Epping NSW 1710, Australia}

\author{D. Campbell-Wilson}
\affil{School of Physics, University of Sydney, NSW 2006, Australia}

\author{J.M. Miller}
\affil{Center for Space Research, Massachusetts Institute of Technology, 70 Vassar Street, 
Cambridge, MA 02139}

\and

\author{M.L.~McCollough}
\affil{Universities Space Research Association, Hunstville, AL 35812}



\begin{abstract}

We report on radio and X-ray observations of the black hole candidate
(BHC) \xte\ performed during its 2000 X-ray outburst.
Observations have been conducted  with the Australia
Telescope Compact Array (ATCA) and have allowed us to sample
the radio behavior of \xte\ in the X-ray Low Hard and
Intermediate/Very High states.
We observed optically thin radio emission from \xte\ five
days after a transition to an Intermediate/Very 
High state, but we observed no radio emission six days later, while 
\xte\ was still in the Intermediate/Very High state.
In the Low Hard state, \xte\ is detected with an inverted radio spectrum.
The radio emission in the Low Hard state most likely originates from a
compact jet; optical observations suggest that the synchrotron emission 
from this jet may extend up to the optical range.  
The total power of the compact jet might therefore be a
significant fraction of the total luminosity of the system. 
We suggest that the optically thin synchrotron synchrotron radio emission detected
five days after the transition to the Intermediate/Very High
state is due to a discrete ejection of relativistic plasma
during the state transition.  Subsequent to the decay of the
optically thin radio emission associated with the state
transition, it seems that in the Intermediate/Very High state
the radio emission is quenched by a factor greater than 50,
implying a suppression of the outflow.  We discuss the
properties of radio emission in the X-ray states of BHCs. 
\end{abstract}



\section{Introduction}

The Soft X-ray Transient (SXT) \xte\ was first detected  by
the All Sky Monitor  (ASM) on-board the {\it Rossi X-ray
Timing Explorer (RXTE)} on  1998, September 7 (Smith 1998).
The outburst was characterized by a ``double peaked'' profile
and a strong (6.8 Crab) and brief (1 day) flare. \xte\ went
through all canonical black hole states (Sobczak et al.\
1999, 2000b; Homan et al.\ 2001) before its return to quiescence in
May 1999.  

Low frequency (0.08-18 Hz) X-ray quasi-periodic oscillations (QPOs) 
as well as a high frequency variable (100-285 Hz) QPO were detected 
during some of the RXTE/PCA observations. This is the fourth  BHC
to display such a high frequency QPO. Based on its strong 
aperiodic variability, QPOs, and X-ray spectrum, \xte\ is believed 
to harbor a black hole, but a mass function has still to be measured 
(Cui et al. 1999; Remillard et al. 1999; Wijnands, Homan, \& van der 
Klis 1999;  Sobczak et al. 2000a and Homan et al. 2001).

Soon after its discovery in X-rays, an optical counterpart was 
reported by Orosz, Bailyn and Jain (1998). A brightening of 4 
magnitudes in V band over the quiescence level was noted by Jain et 
al. (1999). Based on the interstellar absorption lines, a  distance 
of 2.5 kpc  and an optical extinction of 2.2 $\pm$ 0.3 magnitudes 
have been deduced by S{\'a}nchez-Fern{\'a}ndez et al.\ (1999). 
A radio counterpart to \xte\ has been detected at 843 MHz with a 
flux density of $\sim$ 10 mJy by the Molonglo
Observatory Synthesis Telescope (MOST) (Campbell-Wilson et
al. 1998).  Subsequent radio observations indicated that the
strong X-ray flare was accompanied by a large radio flare
with ejection of relativistic plasma, possibly at superluminal
velocities (Hannikainen et al. 2001). 

\xte\ became active in soft X-rays again on 2000 April 2
(MJD 51636; Smith et al.\ 2000).  The Burst and Transient
Source Experiment (BATSE) on-board the {\it Compton Gamma Ray
Observatory (CGRO)} detected \xte\ up to 300 keV (McCollough
et al.\ 2000). Simultaneously, a new optical outburst from
\xte\ was reported (Masetti \& Soria 2000; Jain \& Bailyn
2000).

In this paper we report ATCA observations of the radio
counterpart of \xte, together with daily X-ray observations
performed during the 2000 outburst with RXTE/ASM and
CGRO/BATSE. These observations have been performed in
different X-ray states and we discuss the relation of radio
emission within these states.

\section{Observations and results}

\subsection{Soft and Hard X-ray observations}

The ASM on-board {\it RXTE} (Levine et al.\ 1996) monitors
the activity in soft X-rays of any source in the sky in three
energy bands (1.5--3, 3--5, 5--12~keV).  The ASM light-curve in the
energy band 1.5--12 keV is presented in the top panel of 
Figure~1 and in the lower panel, we plot the hardness ratio
(5--12~keV/3--5~keV) using one-day averages.  The 20--100 keV
light-curve, obtained with BATSE on-board {\it CGRO} using an
Earth occultation technique (Harmon et al.\ 1992), is
presented in the middle panel of Figure~1.  
A power law model with a variable spectral index  has been
folded through the BATSE response matrix for daily flux calculation.
Due to precession of the {\it CGRO} orbit, \xte\ stopped
being occulted on MJD 51682, a few days before the end of the
{\it CGRO} mission.

Figure~1 illustrates the soft and hard X-ray behavior of
\xte\ during the 2000 X-ray outburst.  During the initial rise,
the soft X-ray flux increased slowly, whereas the hard X-ray
flux was already strong when the orbital configuration of
{\it CGRO} allowed its detection by BATSE.  On April 26 (MJD
51660), we note a rapid increase in soft X-rays as well as a
decrease in hard X-rays, which then settled down in a plateau
until the termination of {\it CGRO}.  The decay of this
outburst was a gradual decline in soft X-rays.

Analysis of the RXTE/ASM hardness ratio curve and RXTE/PCA data (Miller et al. in prep)
allowed us to
divide the outburst into three different parts.  The outburst
started and finished with a hard spectrum, whereas the
central part was soft, but with detectable hard X-ray
emission.  This can be understood as follows: an initial
Low Hard state followed by a transition to the Intermediate
state (or Very High state, as this state is now recognized as
a higher count rate version of the Intermediate state, see
Homan et al.\ 2001) on April 26 and then a return to the
Low Hard state after May 13 (MJD 51677).  Preliminary
analysis of RXTE/PCA observations confirm a change in 
power-law photon index from 2.30 to 1.74 between May 16 and
May 24 (Tomsick et al., in prep.).  We note that SXTs are rarely observed
in the Low Hard state in the rising phase of their
outbursts. 

It is interesting to note that this new X-ray outburst occurred
less than a year after the discovery outburst which took
place from 1998 September to 1999 May.  Before that, \xte\
was believed to be in quiescence.  This renewed activity is
unusual for SXTs with long recurrence time (\xte\ probably
belongs to this class, as it has never been detected in the
past years despite sensitive surveys of the Galactic plane).
A parallel can be drawn with \gro\ which also had a new
outburst a few months after the end of series of outbursts
and a period of quiescence (e.g.\ Hameury et al.\ 1997 and 
references therein).  
We should also note that both systems have relatively long orbital 
periods: 2.6 day for \gro\ (Bailyn et al. 1995) and 1.5 day for
\xte\ (Jain et al. 2001). 
The spectral type (F3--F6) of its stellar
companion (Orosz and Bailyn 1997) and  its peculiar position
in the Hertzsprung gap (Kolb et al.\ 1997) may be responsible
for the recurrent outbursts of \gro\ (but see Reg{\"o}s, Tout,
\& Wickramasinghe 1998 for alternative explanations). 
The spectral type of the secondary in \xte\ has not been
identified yet but photometric data are  consistent with a
low mass K0--K5 star  (S{\'a}nchez-Fern{\'a}ndez
et al.\ 1999). Its precise determination would be of great
interest to understand these SXTs with repeated outbursts.
If \xte\ does have a K-type
optical companion, it must be evolved to be large enough  to 
fill its Roche lobe. An evolved companion could conceivably be the similarity
that is responsible  for the repeated outbursts from these systems.

\subsection{Radio observations}

The radio continuum observations of \xte\ were carried out
using the Australia Telescope Compact Array  (ATCA), located
in Narrabri, NSW, Australia. The ATCA is a  synthesis
telescope consisting of 6 antennae, each of 22 m diameter, 
aligned on a 6 km east-west array.  The ATCA radio
observations have been conducted on three  different dates in
2000: April 30, May 6 and June 1. 
The first two observations were performed in the 0.750D compact configuration,
while the last observation was carried out in the high spatial resolution 6B
array. The first and third observations were made at the central frequencies
of 1384, 2496, 4800, and 8640 MHz, with a total bandwidth of 128 MHz; 
observations were made at only 4800 and 8640 MHz during the second observation.

The amplitude and bandpass calibrator was PKS~B1934$-$638 and
the antennae gain and phase calibration was derived from
regular observations (every 20  minutes) of the point source
calibrators B1554$-$64 (at 4800 and 8640 MHz) and B1549$-$790 (at
1384 and 2496 MHz). The editing, calibration, Fourier
transformation, deconvolution and image analysis were
performed using the MIRIAD package (Sault, Teuben, \& Wright
1995). Due to the proximity of a very strong ($\sim$ 145 Jy
at 1 GHz) supernova remnant (G326.3$-$1.8) close to the field
of view, the shortest baselines have not been used in the
analysis of the data at 1384 and 2496 MHz. The ATCA flux
densities of \xte\ are tabulated in Table 1.

The first ATCA observation (on MJD 51665) allows a detection
of the radio counterpart  at a weak level of 7.45 $\pm$ 0.12
mJy and 5.68 $\pm$ 0.06 mJy  at 4800 and 8640 MHz
respectively. For a flux density S$_\nu$ $\propto$
$\nu^\alpha$, a spectral index of $\alpha$ =  --0.46 $\pm$
0.05 is derived. Before this first ATCA observation, the MOST
performed several observations while \xte\ was in its initial
Low Hard state and detected it at a level of 8--15 mJy at
843 MHz.  
The second ATCA observations performed during the
Intermediate/Very High state (on MJD 51670)  did not reveal
significant radio emission from \xte\ with a strong 3
$\sigma$  upper limit of 0.15 mJy at 8640 MHz.

After a transition back to the Low Hard state, we performed
our third ATCA observation  (on MJD 51697) at nearly the end
of the X-ray outburst. Radio emission is  again detected at a
level of $\sim$ 1 mJy at all four frequencies.  The spectrum
is slightly inverted with a spectral index of 0.37 $\pm$ 0.10
(Figure 2a).  The best position of the radio counterpart is:
$\alpha$(J2000) = 15$^h$50$^m$58.7$^s$ and $\delta$(J2000) =
--56\degr 28\arcmin 35.\arcsec 2 with a total uncertainty of
0.\arcsec 3; it is consistent with the position of the optical 
counterpart (Jain et al. 1999).

\section{Discussion}

\subsection{A compact jet in the Low Hard State}

The detection (on June 1st) of a radio source with an
inverted spectrum (spectral index of 0.37 $\pm$ 0.10), while
\xte\ was in the Low Hard state, is reminiscent of the
behavior of the few other BHCs (transient or persistent)
which have been observed at radio frequencies in the Low Hard
state. Indeed, the persistent BHCs from our Galaxy \cx\ 
(Brocksopp et al. 1999), \gx\ (Corbel et al. 2000) and
\grsds\ (Lin et al. 2000)  are found most of the time in the
Low Hard hard state and are detected in radio with a flat
(spectral index $\sim$ 0.0) or inverted spectrum (and also probably \1e). Among the SXTs, we
should note that GS~2023$+$338 (Han \& Hjellming 1992),
GRO~J0422$+$32 (Shrader et al.\ 1994), GS~1354$-$64 
(Brocksopp et al. 2001) and XTE~J1118$+$480 (Hynes et al. 2000, 
Fender et al. 2001a) have been detected with similar
properties (see discussion in Fender 2001b).  \xte\ should now be added to this list. 

A flat or inverted radio
spectrum can be interpreted as  optically thick synchrotron
emission from a compact and conical jet following the model
of  Hjellming \& Johnston (1988). In this model, the higher
frequencies come from the base of the compact jet where the
optical depth is the highest. Low frequency radio emission
arises from the region further from the base. Summing all
contributions over the spatial extent of the jet results in
a flat or inverted spectrum.  Recent VLBA imaging of a
compact jet in the radio core of \cx\ (Stirling et al. 2001)
 confirms this interpretation.  Also the stronger
orbital modulation of the higher frequencies favors this
scenario (Pooley, Fender, \& Brocksopp 1999).  The physical size of these
compact jets is believed to be of the order of $\sim$ 10 a.u at 8.6 GHz.
(corresponding to the m.a.s angular scale).  More VLBA
observations of BHC in the Low Hard state should increase the
number of compact jet detections.

It is expected that the flat spectrum must cut off at high
frequency.  The spectrum of \cx\ is flat up to 220 GHz
(Fender et al.\ 2000a).  But in the case of \gx\ (Corbel \& Fender in prep.)
and XTE~J1118$+$480 (Hynes et al. 2000, Fender et al. 2001a), the inverted
radio spectrum probably extends up to the near-infrared range.
In Figure 2b, we show the
simultaneous radio and optical observations of
\xte\ taken on 2000 June 1 (see Jain et al. 1999, 2001 for a
discussion of the YALO data).  \xte\ is detected at V = 18.9
$\pm$ 0.1 mag. and  I = 16.7 $\pm$ 0.1 mag.
An optical extinction of  A$_\mathrm{V}$ = 2.2 $\pm$ 0.3 mag.  
(S{\'a}nchez-Fern{\'a}ndez et al.\ 1999) has been used  to
deredden these data. Considering the uncertainties in the optical extinction, the spectral
index between the two optical bands (--2.6 $\pm$ 1.0) appears to be non-thermal
and is not compatible with the thermal spectrum of an optically thick 
accretion disk (spectral index $>$ +0.3 in the optical range).
It is therefore possible that a significant fraction of 
the optical-near IR emission  in \xte\ during the Low Hard state
is (optically thin) synchrotron emission from the compact jet.
If this interpretation is correct, the cut-off frequency 
lies in the near infrared range (as in \gx\ and XTE~J1118$+$480).

Using the June 1 observations, the radio luminosity (1--10 GHz band) of the
compact jet  is $\sim$ 7$\times$10$^{28}  \rm \, erg \,s^{-1}$ for a distance
of 2.5 kpc. The luminosity of  the compact jet in
the radio band during the first part of the outburst may be
an  order of magnitude higher, as \xte\ was detected at a level
of 8--15 mJy at 843 MHz  with the MOST.  Assuming that the
inverted spectrum extends to the near infrared ($\sim 10^{14}$ Hz), as discussed above,
we find that the total radiative luminosity of
the compact jet would be of order $\sim$ 2$\times$$10^{34} \rm \, erg \,
s^{-1}$ for the June~1 observation, several per cent of the 2-20
keV band X-ray luminosity of 5$\times$$10^{35} \rm \, erg \,s^{-1}$ 
measured on the same day (Tomsick et al., in prep.). 
Furthermore,  taking account of internal energy of particles within 
the jet and relativistic bulk motion is likely to increase 
the total power required to energize the jet (a precise estimate
would require the knowledge of the inclination angle of the jet). 
Thus, the compact jet is likely very powerful with a total power close to the
bolometric luminosity of the system (accretion disk and corona).

\subsection{Discrete ejection events at a state transition}

The first ATCA observation was performed five days after the
transition from the Low Hard state to the Intermediate/Very
High state. \xte\ is detected around $\sim$ 6 mJy but
contrary to the Low Hard state, the  radio spectrum has a
spectral index of --0.46 $\pm$ 0.05, indicative of optically
thin synchrotron emission.  It is likely that this optically
thin emission is due to a discrete ejection of relativistic
plasma at the time of the state transition when the accretion
disk is unstable.  We would then be detecting the decaying radio
emission from this event and the  emission from
near \xte\ would be low, consistent with our non-detection of
radio emission later in the Intermediate/Very High state. 
During the second part of the 1998-99 X-ray outburst, Homan et
al. (2001) also reported an optically thin synchrotron event after 
a state transition.
Similar discrete ejection events have been inferred during
state transitions in \gx\ (Corbel et al.\ 2000).  It is
possible that almost all X-ray binaries (including neutron
star systems) produce similar weak radio events during state
transitions (see also Hjellming \& Han 1995 and recent work
by Fender \& Kuulkers 2001c).

We note that it was hypothesized that the transition to the
Very High state was accompanied by massive superluminal
ejections (Fender 2000b) and bright flaring radio events.
These new ATCA observations rule out such an interpretation as
no bright ejection event was observed. The formation of these
superluminal ejection events might possibly require a transition
from a quiescent state to a Very High state, i.e. a large and
rapid increase of the mass accretion rate and not a transition from a
state where significant accretion is already occurring (such as the
Low Hard state). Indeed, the discovery outburst of \xte\
was a transition from quiescence to a Very High state.
A bright radio event was detected after the transition and 
possibly resolved in two components moving at apparent
superluminal velocities (Hannikainen et al. 2001).

\subsection{Quenched radio emission in the Intermediate/Very High State}

The second ATCA  observation was performed when \xte\ was in the
Intermediate/Very High state.  At that time, \xte\ was not detected
with ATCA at 4800 and 8640 MHz, with a 3 $\sigma$ upper limit
of 0.15 mJy  at 8640 MHz, i.e. a reduction of radio emission
by a factor greater than 50 if we take into account 
the MOST detection in the initial Low Hard state. This is the
first time radio observations of a BHC have been performed during an 
Intermediate (or Very High) state.
Radio observations in this state are particularly important because 
usually the radio emission is dominated by the decaying optically thin
synchrotron emission associated with the quiescent/active state transition
(and therefore decoupled from the BH/disk system); here, we view the
intrinsic radio properties of the system in this state.

This non-detection is similar to the quenching of the compact jet of \gx\
during the 1998 High Soft state (Fender et al. 1999).
Following the discussion in Fender et al. (1999), it
indicates that  the compact jet is probably physically suppressed in
the  Intermediate state. These new constraints lead us to conclude
that from a radio point of view the Intermediate (or Very
High) and High Soft states are identical. We should note that Belloni 
et al. (1999) classified the 1996 outburst of \cx\ as an Intermediate
state. At that time, radio observations (only performed at the
end of the Intermediate state when \cx\ was returning to its 
standard Low Hard state)  revealed an increase of radio emission
(Zhang et al. 1997), and probably indicating that the radio
emission was indeed lower in the Intermediate state, as observed 
here in \xte.

In the Low Hard state, there is a strong coupling between the
compact jet and the luminosity of the corona (see Brocksopp
et al.\ 1999 for \cx\ and Corbel et al. 2000 for \gx).
Despite the fact that the Intermediate state has significant
hard X-ray emission, it is not a sufficient  condition to
sustain the emission from the compact jet (the power law index 
is different between the Low Hard state and the Intermediate/Very High 
state, but the hard X-ray emission is currently believed to originate
from the corona in both states). Therefore, the Low Hard
state does produces radio emission via a compact jet, whereas the
states with a stronger soft component (High Soft and
Intermediate/Very High) lead to the quenching  of these
compact jets for a reason which is not understood.
Evaporation processes in the standard geometrically thin
disk, below a certain accretion rate, may possibly lead  to
coronal outflows (e.g. Meyer, Liu, \& Meyer-Hofmeister 2000).
It is also possible that high magnetic energy dissipation
might be responsible for the formation of outflows (and
therefore a compact jet above the corona) in the Low Hard
state, whereas the soft states would be dominated by viscous
energy dissipation in the accretion disk  (e.g. Beloborodov
1999; Poutanen 1999; Di Matteo, Celotti, \& Fabian 1999).

\section{Conclusions}

Radio emission has been detected from \xte\ on two occasions during its 2000 outburst. 
Like other BHCs, observed (too rarely) in the Low Hard
state, the radio spectrum of \xte\ is inverted.  This likely
implies that the radio emission is synchrotron emission from
a compact jet. The total power in the 
compact jet may be a significant fraction of the total accretion
luminosity. The Intermediate/Very High state (like the
High Soft state of \gx) is characterized by  a quenching of
this compact jet. It appears that Low Hard states in BHCs
produce compact outflows, whereas the states with a strong
soft component from the accretion disk leads to suppression
of a radio emitting outflow.  State transitions are
accompanied with discrete ejection of relativistic plasma.
More multiwavelength observations of BHCs in the Low Hard
state are needed in order to understand the physics of  these
compact jets.


\acknowledgements 

We thank Drs.\ Ron Ekers, Dave McConnell and the ATCA TAC for
allocating us target-of-opportunity observing time, Steven
Tingay for conducting one of the ATCA observations, and Dr.
George Nicolson for performing HartRAO radio observations of
\xte\ (which resulted in only upper limits due to the proximity 
of the SNR).  S.C. would like to thank Jean-Marie Hameury 
and Richard Hunstead for useful discussions and informations.
The Australia Telescope is funded by the Commonwealth
of Australia for operation as a National Facility managed by
the CSIRO.  RXTE ASM results are kindly provided by the
ASM/RXTE teams at MIT and at the RXTE SOF and GOF at NASA's
GSFC.  SC and PK acknowledge support from NASA grant
NAG5-7405.


\clearpage



\figcaption[Corbel_XTE1550_Fig1.ps1]{Lightcurve of \xte\ using one-day averages. 
Top: RXTE/ASM (1.5--12 keV). Middle: CGRO/BATSE 20--100 keV. Bottom: Hardness 
ratio (5-12/3-5 keV). Arrows indicate the dates of our ATCA radio observations, the dotted 
line the likely period of the transition from the initial Low Hard state (LHS) to the Intermediate/Very High
State (I/VHS). The hatched region shows the approximate date of the return to the final Low Hard state.}

\figcaption[Corbel_XTE1550_Fig2.ps]{Left (a): Radio spectrum on 2000 June 1. The dotted line 
is the best fit function S$_\nu \propto \nu^\alpha$ with $\alpha = +0.37$. Right (b): Radio  and
optical (V,I) measurements of \xte\ on June 1. The optical data originate from 
the Yale 1-m telescope (dereddened using an optical extinction of 2.2 $\pm$ 0.3 mag.).}

\clearpage

\begin{table*}
\begin{center}
TABLE~1

{\sc ATCA observations of \xte.\label{tab_flux}}

\vspace{1mm}
\begin{tabular}{lccccccl}  
\noalign{\smallskip}
\tableline \tableline
\noalign{\smallskip}
              &                      &  & \multicolumn{4}{c}{Flux Density\tablenotemark{a}}     &  \\
\cline{4-7}
\noalign{\smallskip}
Date          & MJD\tablenotemark{b} &  Duration  & 1384 MHz & 2496 MHz & 4800 MHz & 8640 MHz & Spectral index \\ 
              & (days)               &     (hr)   &   (mJy)  &  (mJy)   & (mJy)    &  (mJy)   &    \\
\noalign{\smallskip}
\tableline
\noalign{\smallskip}
\noalign{\smallskip}
2000/04/30 & 51665.07             &    8.0        & \tablenotemark{c} & \tablenotemark{c} & 7.45 $\pm$ 0.12 & 5.68 $\pm$ 0.06 & --0.46 $\pm$ 0.05  \\
2000/05/06    & 51670.96             &    2.5        &   \nodata         &  \nodata          & $<$ 0.10 & $<$ 0.05        & N.A.\\
2000/06/01   & 51697.14             &    5.0        & 0.7 $\pm$ 0.2 & 0.85 $\pm$ 0.09 & 0.88 $\pm$ 0.08 & 1.30 $\pm$ 0.09 & +0.37 $\pm$ 0.10 \\
\noalign{\smallskip}
\tableline
\end{tabular}
\tablenotetext{a}{Upper limits are given at the one sigma level,}
\tablenotetext{b}{Date refers to the beginning of the observation,}
\tablenotetext{c}{No precise flux estimate possible due to the presence of a
bright SNR in the sidelobes.}
\end{center}
\end{table*}

\clearpage

\begin{figure}
\epsfig{file=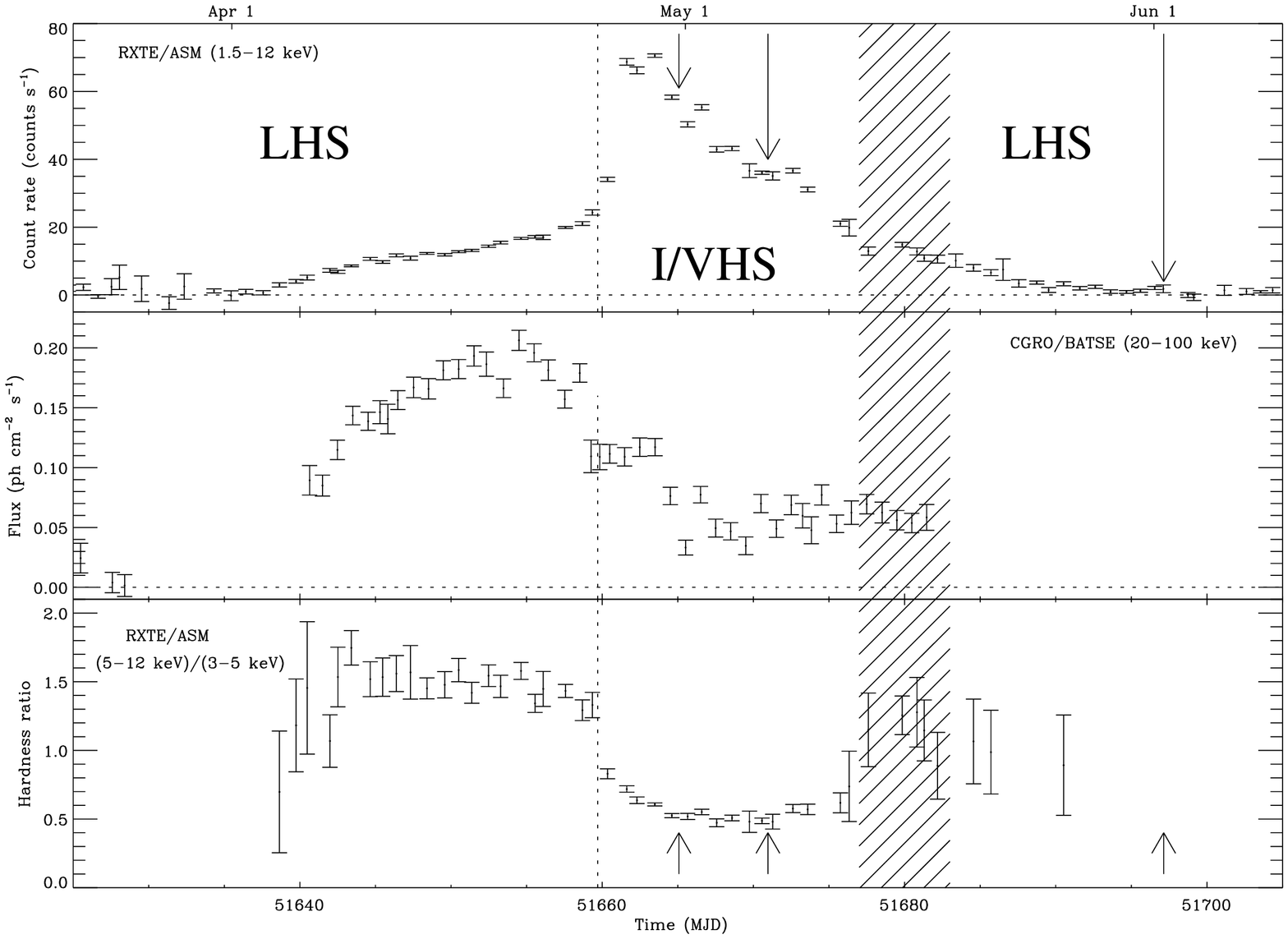,angle=270,width=16cm,clip}
\label{fig1}
\end{figure}

\clearpage

\begin{figure}
\centerline{\hbox{\hspace{2cm}\psfig{figure=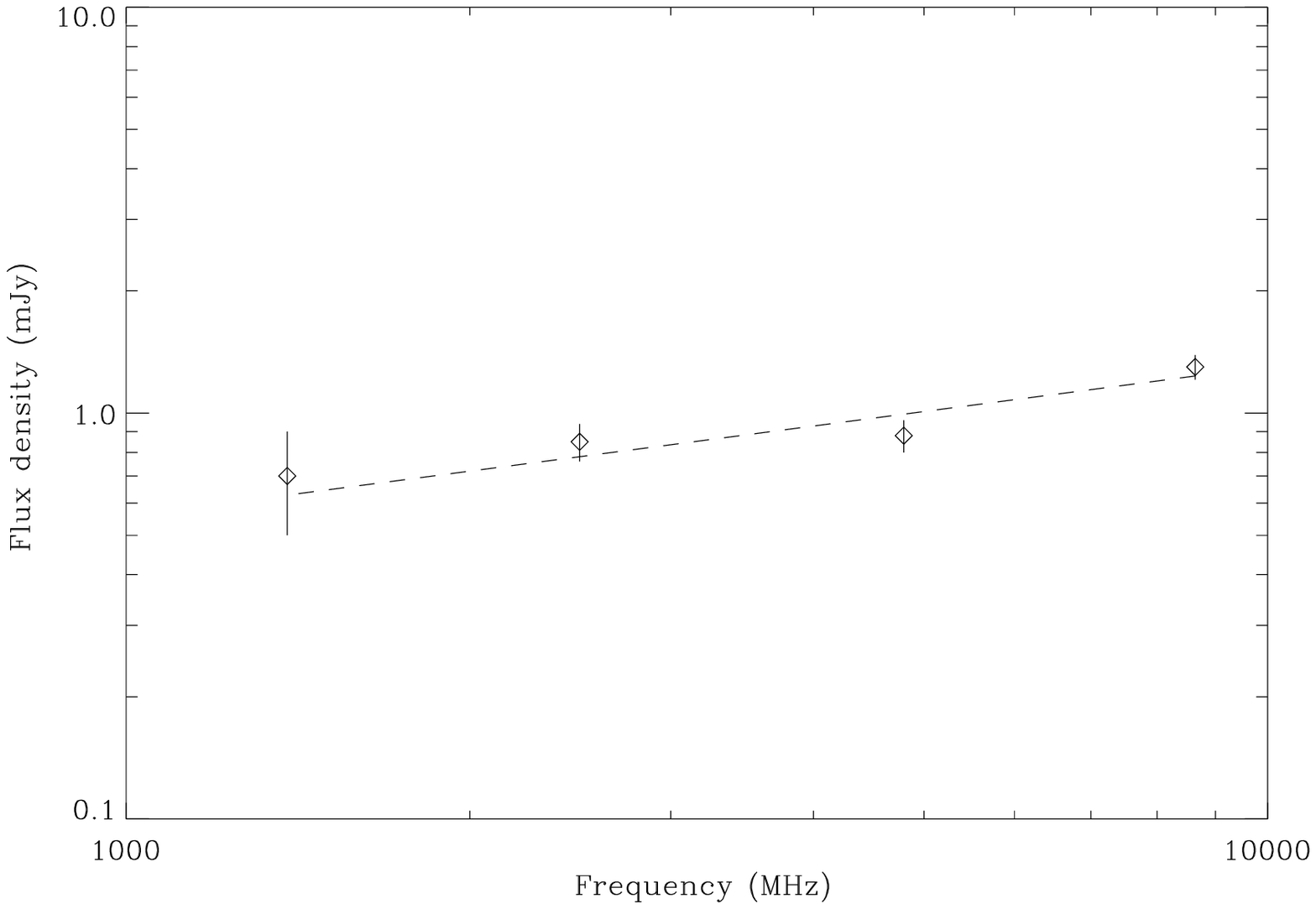,height=7cm,width=8cm,angle=0}\hspace{0cm}
\psfig{figure=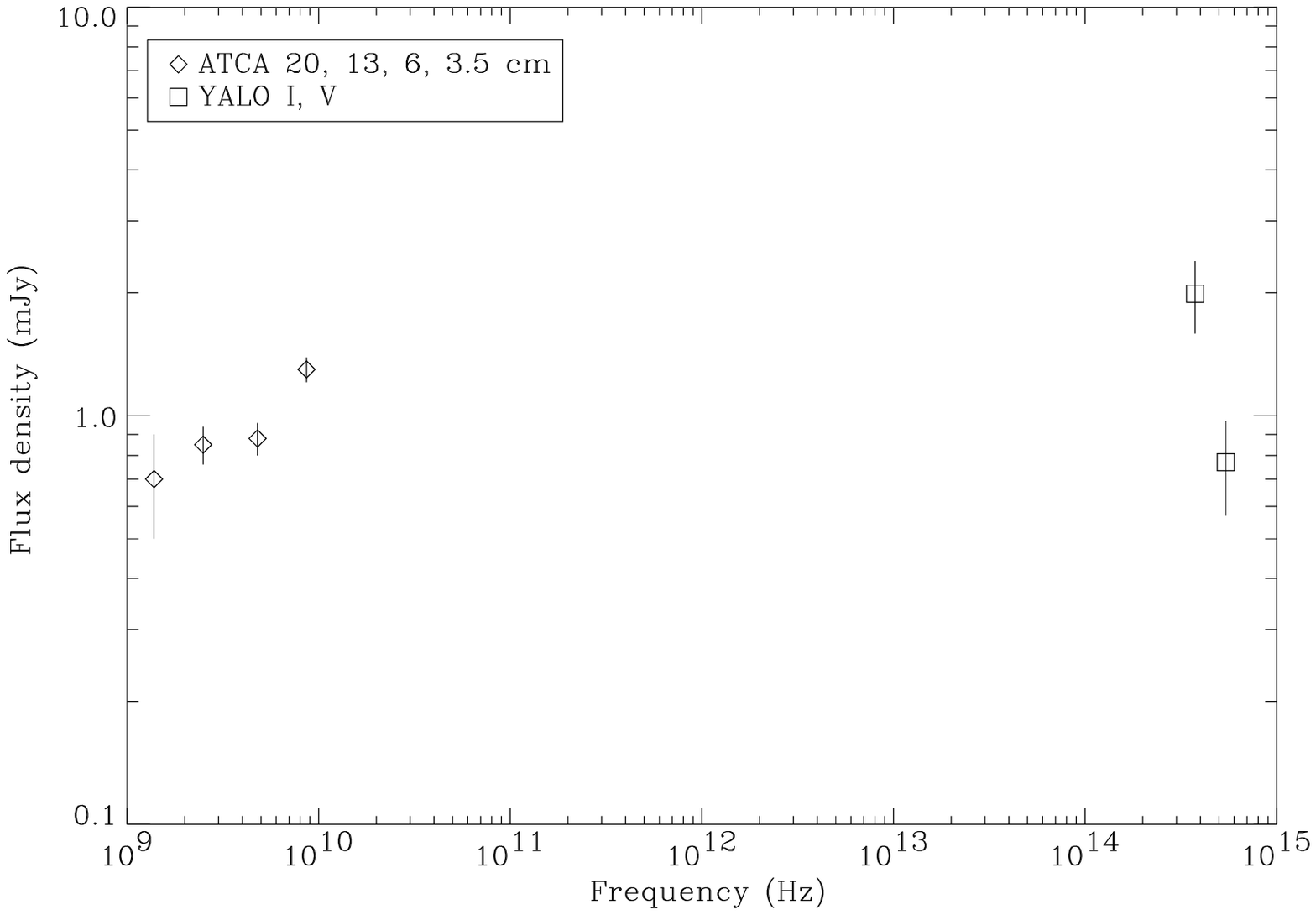,height=7cm,width=8cm,angle=0}}
}
\label{fig2}
\end{figure}

\end{document}